\documentclass[runningheads]{llncs}
\usepackage[T1]{fontenc}
\usepackage{graphicx}
\usepackage{subcaption}
\usepackage{hyperref}
\usepackage{url}
\usepackage{csquotes}
\usepackage{enumitem}
\usepackage{calc}

\usepackage{color}

\usepackage{orcidlink}

\begin{document}
\title{Towards Mixed-Criticality Software Architectures for Centralized HPC Platforms in Software-Defined Vehicles: A Systematic Literature Review\thanks{Preprint submitted to the research paper track of ECSA 2025}}
\titlerunning{Towards Mixed-Criticality Software Architectures}

%
%
\author{Lucas Mauser\inst{1}\orcidlink{0000-0002-9235-5145} \and
Eva Zimmermann\inst{2}\orcidlink{0009-0002-0634-3349} \and
Pavel Nedvědický\inst{2}\orcidlink{0009-0004-4629-8292} \and
Tobias Eisenreich\inst{2}\orcidlink{0009-0004-7168-251X} \and
Moritz Wäschle\inst{1}\orcidlink{0000-0002-7286-1833} \and
Stefan Wagner\inst{2}\orcidlink{0000-0002-5256-8429}
}
\authorrunning{L. Mauser et al.}

\institute{Daimler Truck AG, Leinfelden-Echterdingen, Germany \and
Technical University of Munich, Heilbronn, Germany}
%
\maketitle              
\begin{abstract}
\sloppy Centralized electrical/electronic architectures and High-Performance Computers (HPCs) are redefining automotive software development, challenging traditional microcontroller-based approaches. Ensuring real-time, safety, and scalability in software-defined vehicles necessitates reevaluating how mixed-criticality software is integrated into centralized architectures.
While existing research on automotive SoftWare Architectures (SWAs) is relevant to the industry, it often lacks validation through systematic, empirical methods. To address this gap, we conduct a systematic literature review focusing on automotive mixed-criticality SWAs. Our goal is to provide practitioner-oriented guidelines that assist automotive software architects and developers design centralized, mixed-criticality SWAs based on a rigorous and transparent methodology. First, we set up a systematic review protocol grounded in established guidelines. Second, we apply this protocol to identify relevant studies. Third, we extract key functional domains, constraints, and enabling technologies that drive changes in automotive SWAs, thereby assessing the protocol’s effectiveness. Additionally, we extract techniques, architectural patterns, and design practices for integrating mixed-criticality requirements into HPC-based SWAs, further demonstrating the protocol's applicability. Based on these insights, we propose an exemplary SWA for a microprocessor-based system-on-chip.  
In conclusion, this study provides a structured approach to explore and realize mixed-criticality software integration for next-generation automotive SWAs, offering valuable insights for industry and research applications. 

\end{abstract}
\section{Introduction}\label{sec:introduction}
Today's emerging technologies in automotive, such as powertrain electrification, computing-power-intensive autonomous driving, bandwidth-hungry infotainment systems, enhanced connectivity, and the resulting need for cybersecurity, challenge the component-, domain-oriented development of distributed, embedded Electrical/Electronic (E/E) architectures \cite{Bandur}. Within these, system properties such as busload, computing power, fault rate, modularity, and flexibility are reaching their limits \cite{mauser2024centralization}. The Software-Defined Vehicle (SDV) and its centralized E/E architecture promise to address these limitations as an enabler to accelerate innovation, reduce time-to-market of features, and revolutionize business models \cite{SDV,ExtCom}. It will enable original equipment manufacturers to continuously develop, integrate, and deploy software faster and more flexible \cite{VolvoCICD}. 

Traditional, component-oriented development in the automotive industry resulted in embedded and distributed E/E architectures with over 100 Electronic Control Units (ECUs)~\cite{100ECUs}. By now, there has been an evolution towards partially consolidated, domain-oriented E/E architectures \cite{Bandur}. To differentiate from the competition, key features have been integrated into more powerful domain controllers with the target of developing customer-perceivable application software in-house \cite{100ECUs}. Still, the domain-oriented E/E architecture is characterized by a large number of ECUs widely distributed in the vehicle and hardware-dependent functions, often called logical SoftWare Components (SWCs) \cite{Bandur,mauser2024centralization}. 

Today, automotive E/E architectures evolve towards centralized, zonal E/E architectures, as discussed in our previous work \cite{mauser2024centralization}, \cite{MauserWS}, and by various researchers \cite{HarRed}, \cite{CASEdrivesZonal}, \cite{Dijkstra}, \cite{ParadigmShiftToZonal}, \cite{Bandur}. Centralization is based on functional consolidation into a few, powerful High-Performance Computers (HPCs). Functional decomposition as an exemplary method helps abstract a system and its functionalities to extract SWCs and allocate them to central HPCs \cite{Vogelsang}. This separation of computing and Input/Output (I/O) forms the basis for centralized E/E architectures of SDVs. Here, centralization helps overcome today's distributed system limitations such as busload, computing power, and updateability \cite{mauser2024centralization}. While most logic resides in the HPCs, zonal gateway controllers aggregate distributed, smart I/Os to reduce overall vehicle wiring and weight \cite{HarRed}. More details on different types of E/E architectures can be found in literature \cite{Bandur,HarRed,Zerf,MauserWS}.

\begin{figure}
    \includegraphics[width=\textwidth]{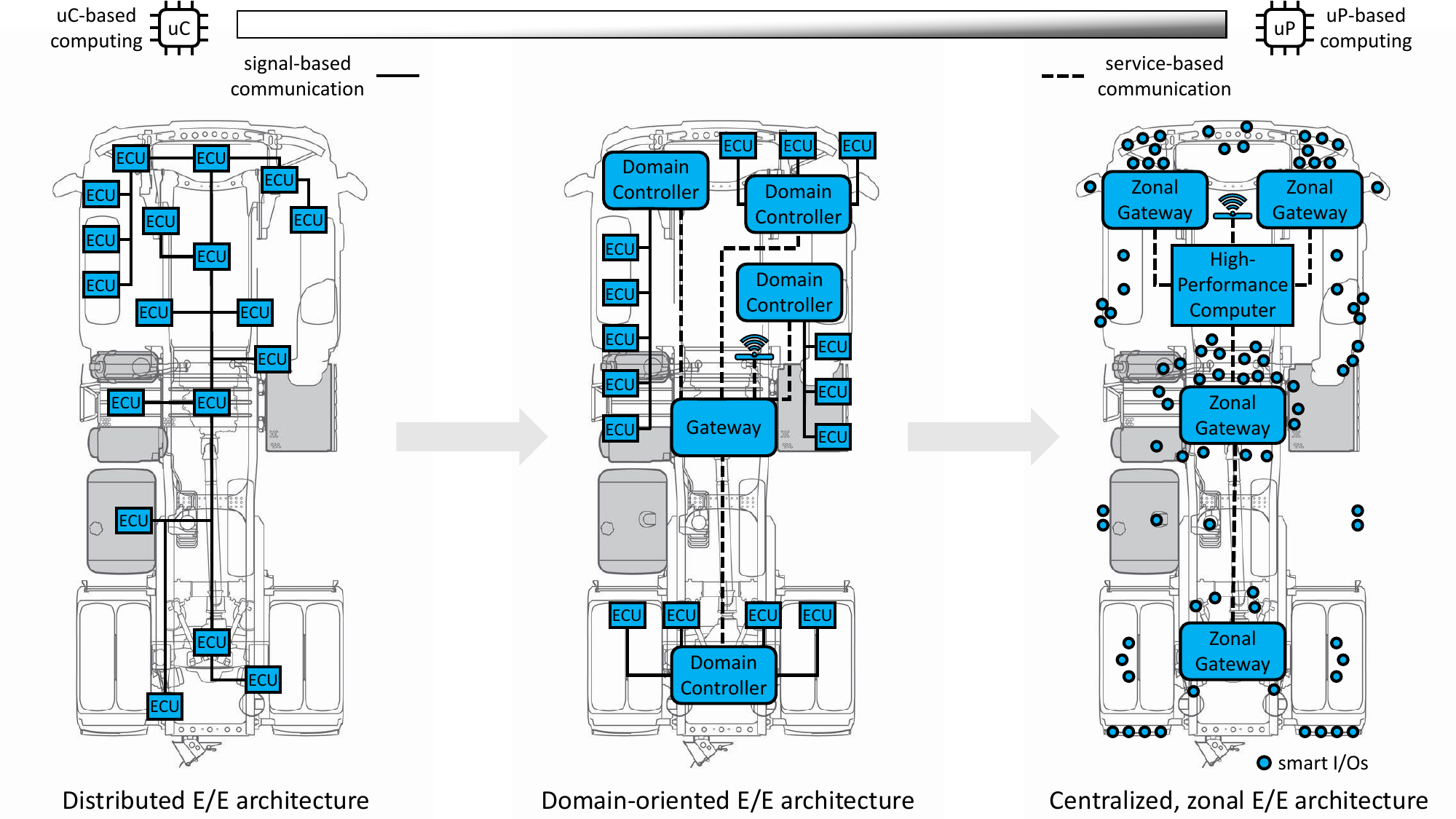}
    \caption{Evolution of automotive E/E architectures} \label{EvolutionOfEEA}
\end{figure}

Functional consolidation within an HPC brings new challenges to its HardWare (HW) and SoftWare (SW) Architecture (SWA). The growing demand for computing power -- driven not only by functional consolidation, but also by emerging technologies such as AI-based algorithms -- is driving a shift from microcontroller-based (\textmu{}C-based) to microprocessor-based (\textmu{}P-based) architectures \cite{ExtCom}. The global development partnership AUTOSAR introduced the \textmu{}C-based platform, AUTOSAR Classic, in 2003, evolving it into a widely adopted SWA for real-time and time-critical applications \cite{autosar-classic}.
The AUTOSAR Adaptive platform was introduced in 2017 to support the aforementioned paradigm shift to meet the demands of service-oriented, high-performance, \textmu{}P-based SWAs \cite{autosar-adaptive}. 
AUTOSAR Adaptive is not intended to replace AUTOSAR Classic. Rather, both platforms will coexist, depending on the specific area of application \cite{Zerf}. In modern vehicles, HPCs integrate consolidated \textmu{}C and \textmu{}P partitions, designed as System-on-Chip (SoC) solutions, offering advantages such as reduced external communication \cite{ExtCom} while backbone networks approach bandwidth limitations \cite{SeedSet1,SeedSet17}. This consolidation of real-time applications with service-oriented, event-driven applications into one single SWA comprises a Mixed-Criticality System (MCS)~\cite{MCS}. MCSs integrate functionalities of different criticality levels. In the automotive context, these are categorized as Automotive Safety Integrity Levels (ASILs) according to the ISO 26262 (QM, ASIL A to D)~\cite{MCS}.  
Meeting the diverse requirements of real-time systems and emerging performance technologies within a centralized SWA leads to our primary research question:\vspace{2mm}

\hypertarget{RQ}{\textbf{RQ:}} \textit{How are automotive, centralized SWAs designed to incorporate mixed-criticality requirements of SDVs?}\vspace{2mm}

Addressing the research question, this paper assists SW architects and developers in consolidating mixed-criticality functions within centralized SWAs of HPC platforms, leveraging the advantages of SDVs. Focusing on technical and methodical enablers, the study bridges academic research and industry practices.

The paper is organized as follows: We review related work to contextualize our research and highlight its added value in Section~\ref{sec:RelWork}. In Section~\ref{sec:RM}, we outline the research method used for the Systematic Literature Review (SLR). The results are presented and discussed in Section~\ref{sec:RD}. Addressing the threats to validity in Section~\ref{sec:TTV}, we conclude and provide an outlook for future research in Section~\ref{sec:CO}.

\section{Related Work}\label{sec:RelWork}
This section reviews related work to position our contributions. While previous studies have examined automotive SWAs, no comprehensive SLR has specifically addressed mixed-criticality architectures for centralized HPC platforms in SDVs.
Aleti et al. \cite{RW-SLRonSWArchOptiMeth} conducted an SLR on SWA optimization methods, presenting a taxonomy to assist architects in selecting techniques. However, since their 2013 study focused on embedded systems, their scope differs from ours, as we propose a centralized, HPC-based SWA rather than optimizing existing ones.
Other SLRs have a narrower focus than our goal of a universal, in-vehicle SWA for centralized HPC platforms. Banijamali et al.~\cite{RW-SLRonCoT} analyzed SWAs for IoT-cloud convergence, discussing attributes, such as portability and maintainability, and Service-Oriented Architectures (SOAs) -- key considerations for SDVs \cite{MauserWS,mauser2024centralization}. However, it does not specifically address in-vehicle architectures. Similarly, Avci et al.~\cite{RW-SLRonBigData} studied big-data architectures across industries, identifying automotive use cases offloaded to the cloud rather than in-vehicle architectures, as explored in our research \cite{mauser2024centralization}.
Beyond SLRs, several Systematic Mapping Studies (SMSs) provide broader overviews of automotive SWA research. 
However, their general scope limits their relevance to the specific challenges of centralized HPC platforms in SDVs.
For example, Haghighatkhah et al.~\cite{RW-SMSAutoSWEng} analyzed literature on Automotive Software Engineering (ASE), identifying \enquote{system/software architecture and design, qualification testing, and reuse} as the three most frequently studied areas. Their findings highlight the critical role of SWA and design in academia and industry, as evidenced by the leading ASE research institutions spanning companies and public organizations. The review highlights research with high industrial relevance but low scientific rigor. Here, we derive and conclude, in alignment with the conclusion of Haghighatkhah et al., the need for practitioner-oriented \enquote{guidelines for selecting existing solutions, technologies, and practices} based on a rigorous SLR satisfying scientific expectations.    


\section{Research Method}\label{sec:RM}
We conducted an SLR to answer the research question to address the identified need for systematic research. This section highlights the key milestones of the three phases of planning, conducting, and reporting to ensure the key characteristics of a credible SLR according to Kitchenham et al.~\cite{kitchenhamguidelines}: comprehensiveness, reproducibility, and transparency. The working products of these key milestones are documented and publicly available on \href{https://github.com/AutomotiveArchitectures/ECSA-2025}{GitHub} and \href{https://doi.org/10.5281/zenodo.15093879}{Zenodo}.  



\subsection{Research Question}
Kitchenham et al.~\cite{kitchenhamguidelines} emphasize the importance of the research question as the essential part of a systematic review. Thus, we applied the PICOC (Population, Intervention, Comparison, Outcomes, Context) criteria presented within the guidelines to refine and reframe the research question. The PICOC criteria determine the research scope, ensure the consideration of different viewpoints, and help reduce bias.
\begin{flushleft}
\textbf{P:} Automotive SWAs -- \textbf{I:} Centralized HPC SWA -- \textbf{C:} Real-time, embedded SWA -- \textbf{O:} Mixed criticality SWAs -- \textbf{C:} Technological transformation in automotive industry.
\end{flushleft}
With these, we substantiate our research question: \textit{How are automotive, centralized SWAs designed to incorporate mixed-criticality requirements of SDVs?}


\subsection{Review Protocol}\label{sec:RP}
Three performance indicators guide the selection of our SLR protocol: recall, precision, and F-measure. Recall ensures the review's comprehensiveness by minimizing the risk of missing relevant studies, while precision ensures identified studies contribute to answering the research question. The F-measure, the harmonic mean of precision and recall, balances both metrics. \cite{hybrid1} 
For this study, we selected an SLR based on a Scopus database search, which has demonstrated high precision in evaluations by Mourão et al. \cite{hybrid1}. This approach prioritizes relevance over recall, avoiding the lower precision seen in databases like Google Scholar \cite{hybrid1}. 
Threats to validity of the protocol will be discussed in Section~\ref{sec:TTV}.

\subsection{Search String}
For the Scopus search, we constructed the following search string based on the PICOC criteria:
\begin{flushleft}
TITLE-ABS-KEY ("software architecture*" AND ( vehicle OR automotive ) AND ( "mixed critical*" OR hpc OR "High*Performance Comput*" OR "central*" OR "functional domain*" ))
\end{flushleft}

The search string is applied to titles, abstracts, and keywords in Scopus (TITLE-ABS-KEY). We discussed, reviewed, and modified the search string in review sessions with all authors to ensure plausibility and reduce overlooks of a single author. The search string is intended to answer the research question by identifying modern, mixed-criticality automotive SWAs in recent literature. 

\subsection{Inclusion and Exclusion Criteria}
To produce the seed set of studies from the database search, we defined the inclusion and exclusion criteria in Table~\ref{tab1}. These include content-related criteria and quality-related criteria for achieving a certain level of quality, as discussed in the guidelines of Kitchenham et al. \cite{kitchenhamguidelines}. For a study to be included, content-related and quality-related criteria must be valid.
The exclusion of non-peer-reviewed publications ensures the inclusion of high-quality studies. For the same reasons, we exclude books or sections from books, as peer review is not guaranteed. Based on the identified literature and the evolution of automotive E/E architectures described in Section~\ref{sec:introduction}, the SDV approach, which separates computing and I/O, gained momentum in the mid-2010s. This can also be confirmed by analyzing the search results by publication year. As a result, we exclude studies from the HW-driven and ECU-oriented era of E/E architectures before 2010.
The study language is limited to English and German, in which the authors are proficient.

\begin{table}[h]
\caption{Inclusion and exclusion criteria}\label{tab1}
\begin{tabular}{|p{0.49\textwidth}|p{0.49\textwidth}|}
\hline
A study is included if & A study is excluded if\\
\hline
\parbox[t]{0.49\textwidth}{\raggedright \textbf{IC1}: the title, abstract, and keywords clearly demonstrate a contribution to the research question\\ \textbf{IC2}: the abstract tends to answer the aspects of data extraction derived from the research question} & \parbox[t]{0.49\textwidth}{\raggedright \textbf{EC1}: the study does not meet any of the inclusion criteria\\ \textbf{EC2}: the study is not peer-reviewed\\ \textbf{EC3}: the study is either a book or a section from a book\\ \textbf{EC4}: the study is published before 2010 or after December 2024\\ \textbf{EC5}: the study is not written in English or German\\ \textbf{EC6}: the study is not accessible}\\
\hline
\end{tabular}
\end{table}

We use the tool Rayyan to efficiently document and perform the inclusion/ex\-clusion process \cite{Rayyan}. Each study’s inclusion and exclusion criteria are documented, with references to the short names of the corresponding criteria and a detailed rationale.
The first four authors reviewed each study from the Scopus database search. This process served as a pilot to establish a common understanding and consensus on the inclusion/exclusion criteria while reducing the potential for bias from a particular author. Any initial disagreements, discussions, and resolutions have been documented in the SLR report.

\subsection{Data Extraction}
For data extraction of the included papers, we broke down the \hyperlink{RQ}{research question} into Data Extraction Criteria (DEC) to further explore the particular aspects. The criteria facilitate data extraction and the subsequent synthesis of the study outcomes.

\begin{enumerate}[label=DEC\arabic*,leftmargin=\widthof{DEC1~~},labelsep=\widthof{~~}]
    \item \label{ExtrctnC1} Which functional domain(s) does the study analyze and/or modify in relation to SWA changes?
    \item \label{ExtrctnC2} Which constraint(s) does the study identify as drivers for SWA changes?
    \item \label{ExtrctnC3} Which technologies does the study identify as enablers or catalysts for changes in the SWA?
    \item \label{ExtrctnC4} How does the study technically address the integration of diverse SW requirements (real-time, non-real-time, safety-critical, etc.) within a centralized automotive SWA?
    \item \label{ExtrctnC5} Which architectural patterns or design practices are proposed to systematically support mixed-criticality in centralized automotive SWAs?\vspace{2mm}
\end{enumerate}

To effectively divide the work of data extraction while minimizing bias, only the first
author reviewed all of the literature in the seed set. Authors 2, 3, and 4 each reviewed one-third of the seed set. The extracted data were compared between the first author and the other reviewers. Agreement on the results was required, and any ambiguities have been resolved.

\subsection{Data Synthesis}

The data synthesis is designed based on the extraction criteria. \ref{ExtrctnC1}, \ref{ExtrctnC2}, and \ref{ExtrctnC3} provide context on why SWAs are facing change. For \ref{ExtrctnC1}, we conducted a quantitative, deductive coding to highlight functional domains according to the domain taxonomy Vehicle Signal Specification (VSS) V5.0 by the COVESA community \cite{VSS}. For \ref{ExtrctnC2} and \ref{ExtrctnC3}, we conducted an explorative, qualitative open coding by extracting and grouping identified constraints and technologies from the literature without applying weightings.
\ref{ExtrctnC4} and \ref{ExtrctnC5} answer the \hyperlink{RQ}{research question}. We conducted a qualitative content analysis using open coding for both: \ref{ExtrctnC4} to discuss technical implementations and approaches, and \ref{ExtrctnC5} to examine architectural patterns and design practices. The results of the data extraction and synthesis are presented and discussed in the following section.

\section{Results and Discussion}\label{sec:RD}
The search string applied in the database Scopus resulted in 97 identified studies. The inclusion/exclusion criteria application resulted in 21 studies included in the final set. The data extraction of these 21 studies is the basis for the results in this section. While we discuss each of the included studies below, we list them here in a preliminary manner to provide a central overview:\\ \cite{SeedSet1,SeedSet2,SeedSet3,SeedSet4,SeedSet5,SeedSet6,SeedSet7,SeedSet8,SeedSet9,SeedSet10,SeedSet11,SeedSet12,SeedSet13,SeedSet14,SeedSet15,SeedSet16,SeedSet17,SeedSet18,SeedSet19,SeedSet20,SeedSet21}.

\subsection{Providing context}
\begin{figure}[h!]
    \centering
    \captionsetup{justification=centering}
        \includegraphics[width=\textwidth]{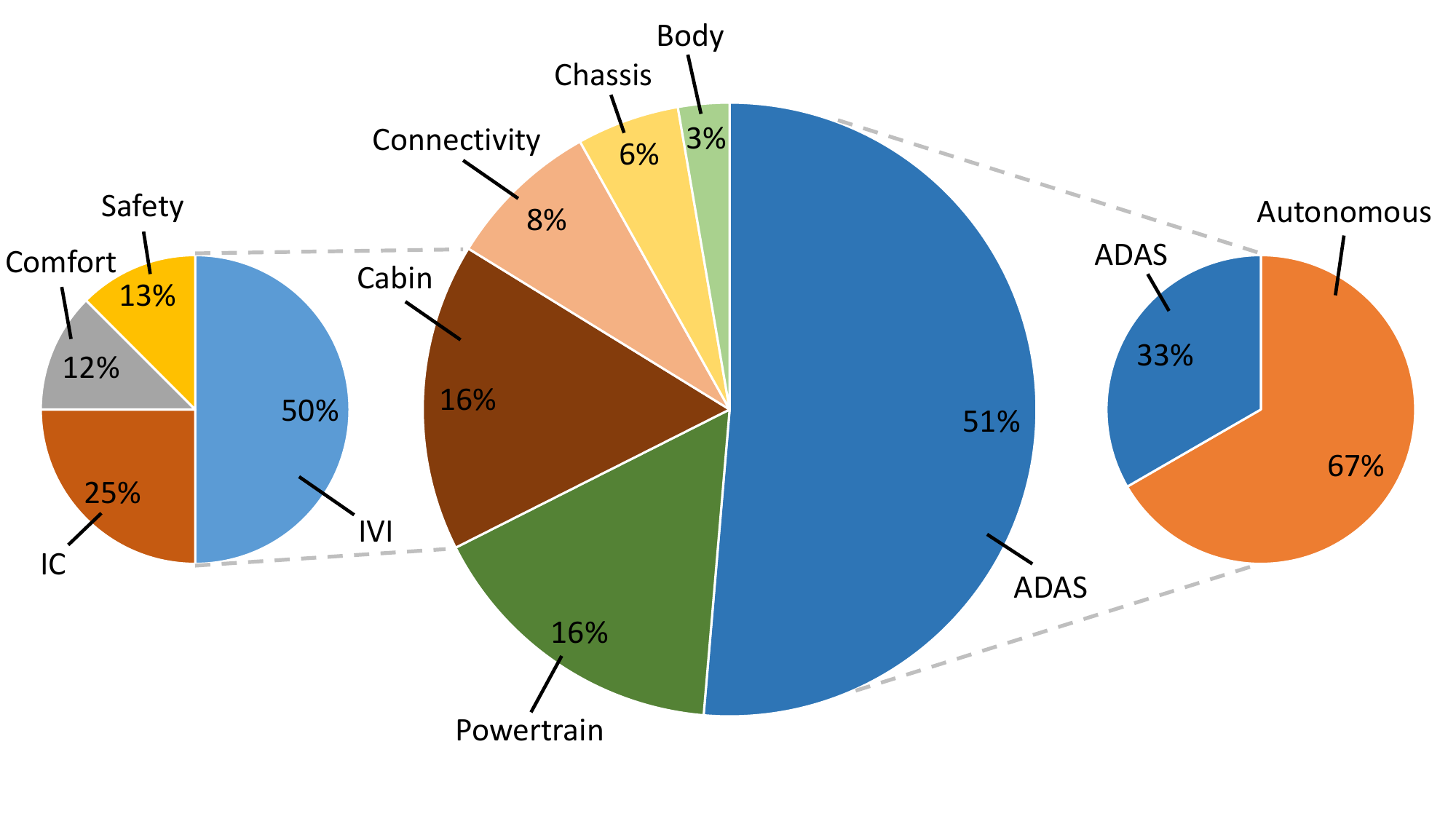}
        \caption{\protect\ref{ExtrctnC1} - Functional domains driving change in SWAs according to Vehicle Signal Specification (VSS) V5.0 \cite{VSS}} 
    \label{EC1}
\end{figure}

Fig.~\ref{EC1} illustrates the key functional domains driving change in automotive SWAs. The three most influential domains are: Advanced Driver Assistance Systems (ADASs), particularly its autonomous subcategory; powertrain, driven by electrification; and cabin, driven by infotainment.
ADASs and autonomous vehicles demand high computing power to support computer vision algorithms, pushing embedded, \textmu{}C-based ECUs and their SWAs to their limits \cite{SeedSet12}.
In the powertrain domain, electrification expands the design space, increasing the complexity of SW implementations and necessitating a fundamental rethinking of embedded SWAs \cite{SeedSet6}.
The evolution of high-performance digital displays challenges Instrument Cluster (IC) and In-Vehicle Infotainment (IVI) systems, not only by increasing HW and SW complexity but also by requiring integration of safety-critical and non-safety-critical applications within a single, centralized SWA \cite{SeedSet2}.

\begin{figure}[h!]
    \centering
    \begin{subfigure}{0.49\textwidth}
        \centering
        \includegraphics[width=\textwidth]{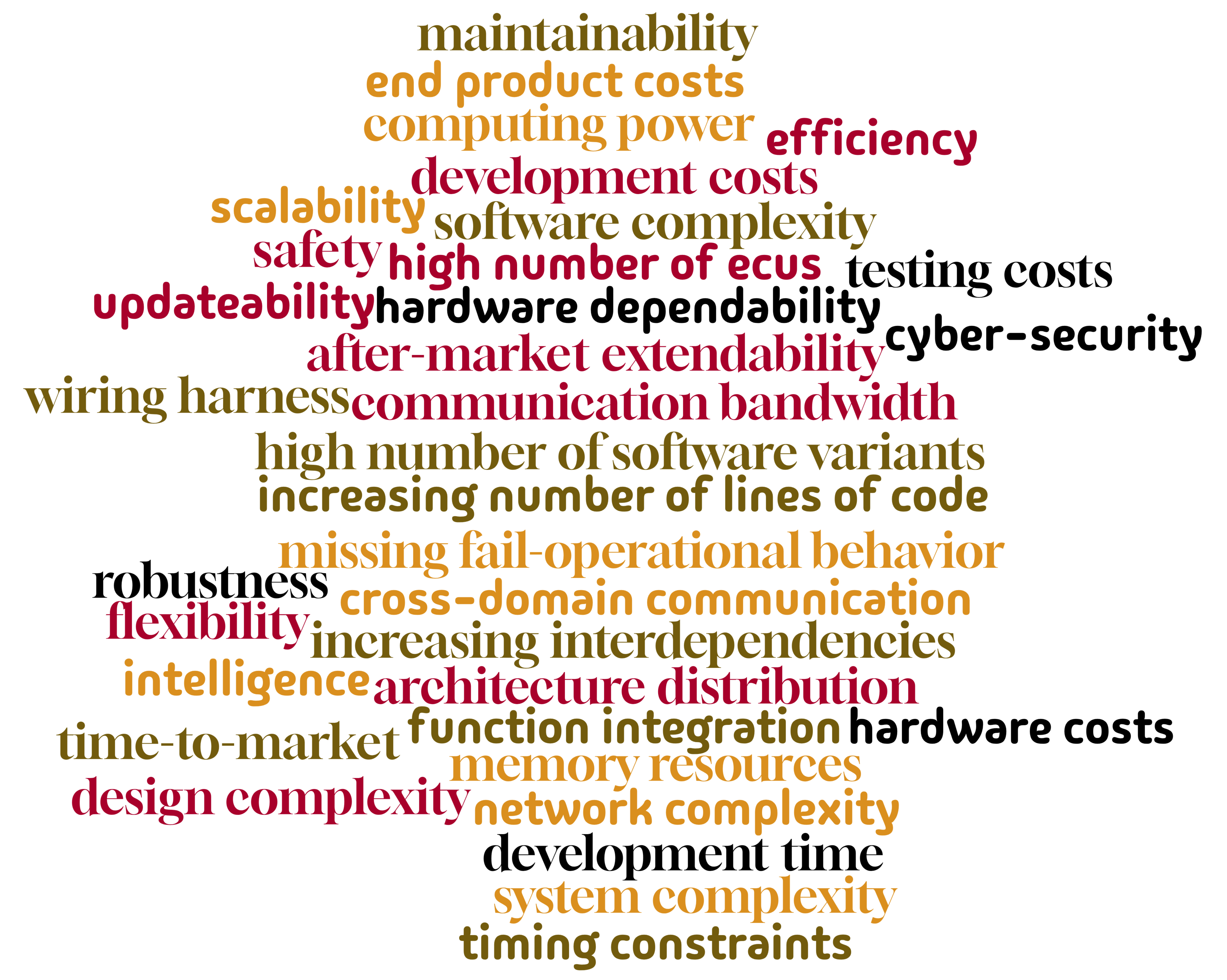}
        \caption{Constraints driving change}
        \label{fig:EC2}
    \end{subfigure}
    \hfill
    \begin{subfigure}{0.49\textwidth}
        \centering
        \includegraphics[width=\textwidth]{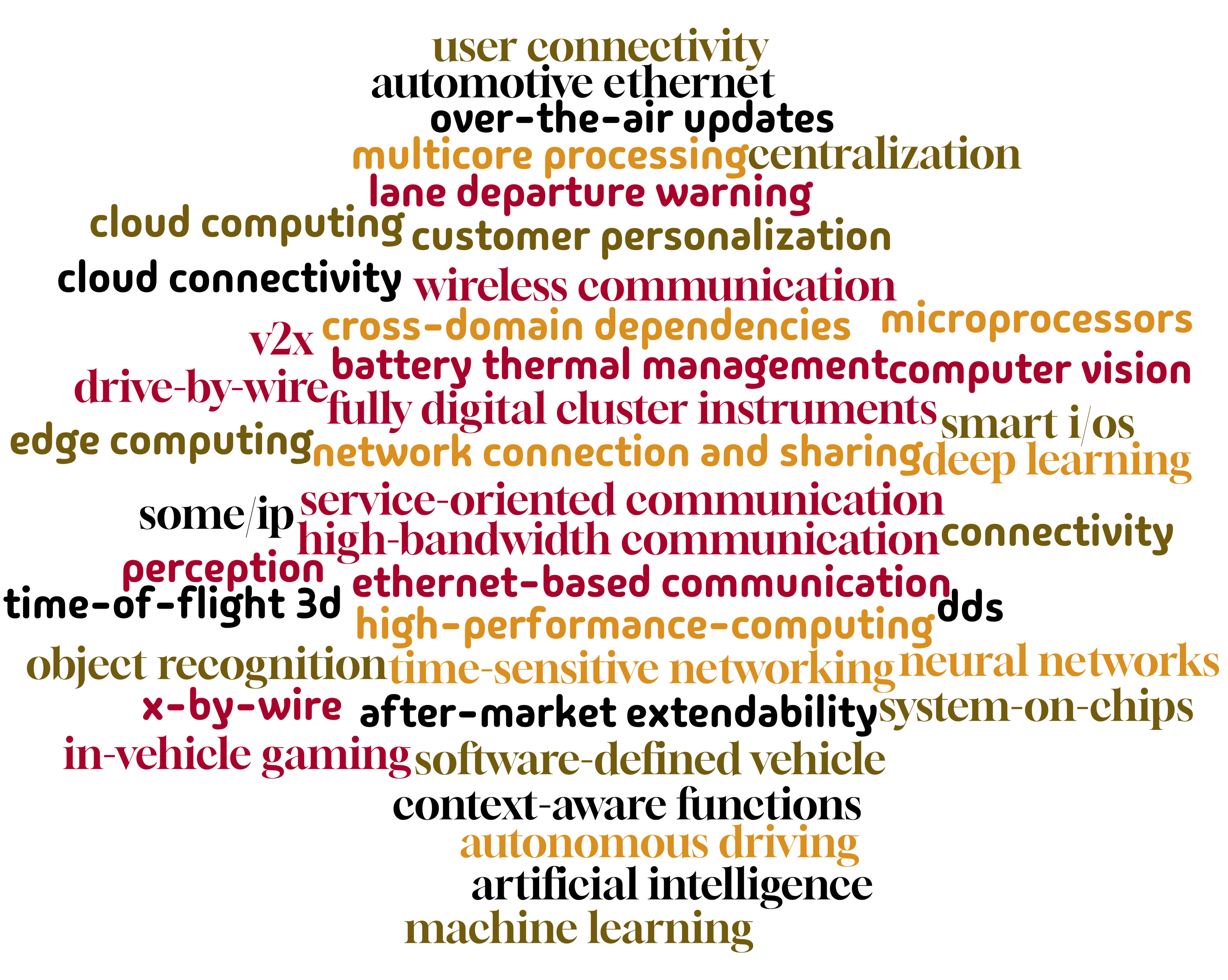}
        \caption{Technologies driving change}
        \label{fig:EC3}
    \end{subfigure}
    \caption{Word clouds \protect\ref{ExtrctnC2} (left) and \protect\ref{ExtrctnC3} (right)}
    \label{EC2EC3}
\end{figure}

Such constraints of today's SWAs, identified in the literature by \ref{ExtrctnC2}, and technologies as enablers and catalysts for changes in SWAs, identified in the literature by \ref{ExtrctnC3}, are illustrated by the word clouds in Fig.~\ref{EC2EC3}. 

With these extracted and synthesized data providing context, we want to focus on the \hyperlink{RQ}{research question} of Section~\ref{sec:introduction}. To answer the research question, we set up \ref{ExtrctnC4} and \ref{ExtrctnC5}, the results of which we will discuss qualitatively below.

\subsection{Answering the research question}
\begin{figure}
    \centering
    \includegraphics[width=\columnwidth]{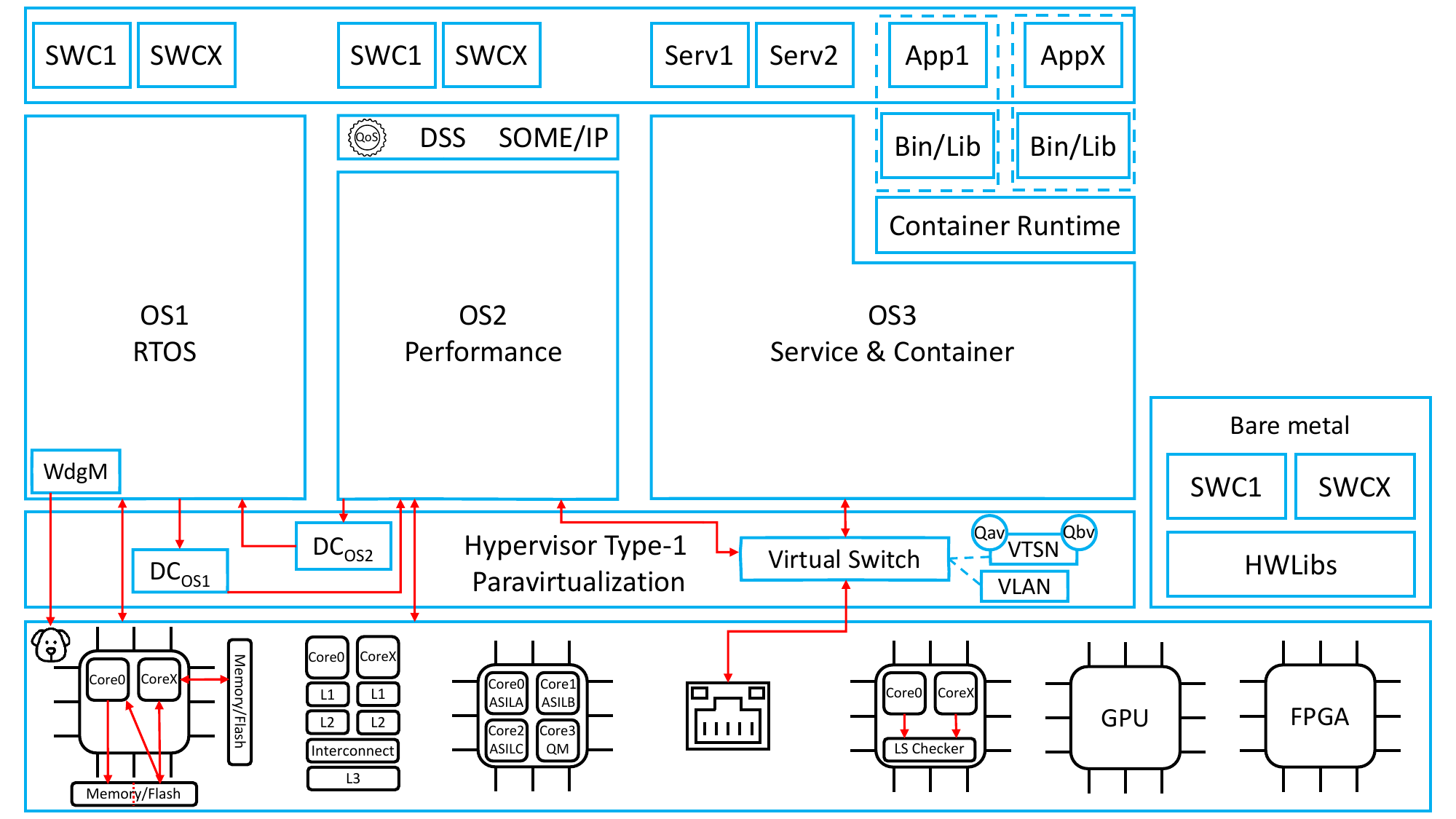}
    \caption{SoC-based, mixed-criticality SWA}
    \label{fig:ExemplarySWA}
\end{figure}

We analyze \ref{ExtrctnC4} and \ref{ExtrctnC5} inspired by the AUTOSAR structure and its layers, starting from the HW layer, through HW-near SW layers, the operating system layer, the middleware layer as so-called basic SW in AUTOSAR Classic, to the application SW layer \cite{autosar-classic,autosar-adaptive}. Based on the following discussion, we integrated the findings into the exemplary SWA shown in Fig.~\ref{fig:ExemplarySWA} illustrating key aspects for realizing mixed-criticality SWAs for HPC platforms. Details and the derivation will be elaborated in the following, based on the analysis of \ref{ExtrctnC4} and \ref{ExtrctnC5}. 

\vspace{1mm}

\textbf{HW Layer}: High-performance SoC HW forms the foundation of centralized HPC platforms \cite{SeedSet2,SeedSet16,SeedSet18}. Literature suggests multi-core and SoC designs to allocate SWCs with similar properties, such as criticality class, to the same core or SoC semiconductor partition \cite{SeedSet6,SeedSet8,SeedSet12}. Placing dedicated SWC sets on individual cores facilitates space partitioning and ensures Freedom from Interference (FFI)~\cite{SeedSet4,SeedSet17,SeedSet18}. For highly parallel workloads, many-core processors optimize specific computational aspects, often at the expense of other performance trade-offs \cite{SeedSet14}. Similarly, Field-Programmable Gate Arrays (FPGAs) are increasingly used in automotive HPC platforms as energy-efficient, reconfigurable HW accelerators with low-latency processing \cite{SeedSet4}. Private memory and cache HW partitions improve timing predictability and FFI, but at the same limit parallelization and timing gains due to static (design-time) core allocation in comparison to dynamic (run-time) core allocation \cite{SeedSet5,SeedSet14,SeedSet18}. Mode-based dynamic core allocation can further improve performance by adapting to varying criticality levels and environmental changes in real-time, ensuring optimal resource utilization and scalability for ADASs \cite{SeedSet8,SeedSet18}. To improve fault tolerance, lock-step operation modes distributed across available cores can detect failures in HW components and increase system reliability \cite{SeedSet21}. Redundant flash memory partitions are identified to enhance the system's robustness and consistency \cite{SeedSet13}. Fig.~\ref{fig:ExemplarySWA} illustrates these capabilities for extensive resource partitioning and redundancy approaches at HW layer as the foundation for a centralized, mixed-criticality SWA.

\textbf{HW-near SW Layer}: At the HW-near SW layer, HW resources must be allocated and managed efficiently. The literature suggests virtualization via hypervisors as a key approach \cite{SeedSet2,SeedSet8,SeedSet16,SeedSet18,SeedSet19}. Hypervisors enable the creation of Virtual Machines (VMs), each running a required Operating System (OS), or even bare-metal applications \cite{SeedSet2,SeedSet4}. There are two commonly used types of hypervisors. Type 1 (bare-metal) hypervisors run directly on the HW, providing exclusive and static resource management, leading to higher efficiency, performance, and strongest FFI. Type 2 hypervisors run on a host OS, making HW access less efficient and introducing a risk of failure if the host OS crashes \cite{SeedSet4,SeedSet18,SeedSet19}. Additionally, virtualization can be classified as full or paravirtualization. Full virtualization emulates HW components, allowing unmodified guest OSs to run as if on physical HW. Paravirtualization requires guest OS modifications to interact more efficiently with the hypervisor, improving performance. The literature confirms paravirtualization offers higher performance than full virtualization, which incurs overhead due to HW emulation \cite{SeedSet4,SeedSet19}. Virtualization is also seen as having \enquote{cost, reliability, availability, and adaptability} benefits while guaranteeing spatial and temporal isolation \cite{SeedSet4,SeedSet8}. In summary, virtualization enables the separation of SW logic from HW and mixed OS environments as a prerequisite for SDVs \cite{SeedSet15}. 

\textbf{OS Layer}: At the OS layer, Ferraro et al. \cite{SeedSet4} propose a Real-Time OS (RTOS) for actuation and a Linux OS for performance-intensive object detection. Additionally, Robot Operating System 2 (ROS2), based on Data Distribution Service (DDS), is used in self-driving applications to enhance scalability \cite{SeedSet4}. Similarly, Lee and Wang \cite{SeedSet15} split sensing/actuation-heavy SW and processing-heavy SW. Niedballa and Reuss \cite{SeedSet13} employ an RTOS for safety- and time-critical tasks and system monitoring, while Linux OS handles computing-intensive application logic.

\textbf{Middleware Layer}: At the middleware layer, various methods and protocols support the coexistence of mixed-criticality OSs and applications. With increasing cross-domain communication \cite{SeedSet1,SeedSet17} and dependencies \cite{SeedSet12}, a key challenge is ensuring FFI as centralized HPC platforms consolidate multiple domains \cite{SeedSet1,SeedSet17}. Holstein and Wietzke \cite{SeedSet2} address this risk for inter-VM communication by introducing architectural approaches such as the \textit{clear separation approach}, \textit{layers of interconnections}, and \textit{minimalistic approach} with its one-way/read-only Data Containers (DCs) illustrated in Fig.~\ref{fig:ExemplarySWA}. To maintain data integrity in mixed-criticality environments, key enablers include memory management and stack monitoring mechanisms such as isolated communication channels, access control mechanisms, signed communication data, consistency checks, redundant memory, multiple memory copies, and checksums \cite{SeedSet2,SeedSet5,SeedSet13}.
For both communication and isolation, a Virtual Local Area Network (VLAN) enables FFI by creating virtualized network segments \cite{SeedSet4}. Also, Virtual Time-Sensitive Networking (VTSN) extends Quality-of-Service (QoS) capabilities to VMs using TSN schedulers Qav (bandwidth reservation based on criticality or priority) and Qbv (communication separation into fixed-duration periodic cycles)~\cite{SeedSet4}. Although Qbv achieves lower end-to-end latency, Ferraro et al.\cite{SeedSet4} suggest combining both approaches to improve reliability and robustness. Ensuring that both functional and non-functional requirements are met at runtime is critical to meeting QoS requirements \cite{SeedSet10}. Similarly, DDS is proposed as middleware to ensure \enquote{reliable, scalable, and efficient real-time communication} \cite{SeedSet9}. Additionally, time partitioning can enhance predictability and reliability in program execution, which Watchdog Managers (WdgMs) can monitor \cite{SeedSet6,SeedSet17}. Furthermore, self-diagnostic capabilities can enhance software-sided redundancy, enabling hot-switching between redundant components to mitigate failures and maintain system continuity \cite{SeedSet11}.
\sloppy In general, the literature identifies a shift from statically configured, signal-oriented communication towards a more flexible, scalable, and updateable service-oriented communication \cite{SeedSet1,SeedSet3,SeedSet17}. This evolution is exemplified by the publisher-subscriber mechanism implemented in the automotive communication protocol Scalable service-Oriented MiddlewarE over IP (SOME/IP) \cite{SeedSet3,SeedSet15}, which decouples communication between applications \cite{SeedSet17}. However, this approach introduces timing uncertainties, which can be managed using the previously mentioned mechanisms, such as TSN, DDS, and other timing-aware solutions \cite{SeedSet3}. This evolution lays the foundation for centralized, data-centric Automotive Service-Oriented Architectures (ASOAs), enabling the SDV approach to update vehicles post-production in a more flexible manner and thus to evolve over time in response to new customer needs, while also reducing time to market \cite{SeedSet4,SeedSet9,SeedSet10,SeedSet18}. Several frameworks support these advancements at the middleware layer, including: SOAFEE \cite{SeedSet2} providing real-time guarantees in SOAs by integrating cloud-native practices with automotive requirements; Time-Sensitive Autonomous Architecture (TSAA)~\cite{SeedSet4} leveraging TSN protocols to support mixed-criticality SDVs; Brain Centralized Electronic and Electrical Architecture (BCEA)~\cite{SeedSet16} enabling HW-SW decoupling, facilitating offloading to cloud computing; Hypervisor-based Fault tolerance approach for heterogeneous Automotive Real-time systems (HyFAR)~\cite{SeedSet19} enabling mixed-criticality via virtualization technology; and AUTOSAR Adaptive as a middleware (see Sec.~\ref{sec:introduction}) introducing ASOAs with enhanced updateability \cite{SeedSet14,SeedSet19} for centralized HPCs with high computing power and high-speed communication demand \cite{SeedSet16}.

\textbf{Application SW Layer}: At the application SW layer, Lee and Wang \cite{SeedSet15} advocate for microservice architectures to enhance evolvability and accelerate software development and release, addressing the growing complexity of automotive software \cite{SeedSet7}. This aligns with broader recommendations in the literature to design loosely coupled SWCs that maximize HW independency, improving transferability and scalability \cite{SeedSet15,SeedSet16,SeedSet17}. Li et al. \cite{SeedSet16} highlight HW-SW decoupling as a critical enabler for cloud computing, facilitating computational offloading while maintaining compatibility with HPC-based architectures. To further enhance portability, isolation, and deployment flexibility, containerization is proposed as a lightweight alternative to full virtualization. In automotive HPC platforms, containers are typically deployed atop a hypervisor, combining the strong isolation of VMs with the efficiency of containers. This setup enables multiple guest OSs to run separate containerized applications, ensuring scalability, modularity, and rapid software updates while preserving FFI between mixed-criticality applications \cite{SeedSet3}. Additionally, Kugele et al. \cite{SeedSet20} propose architectural patterns, such as the \textit{User Shadow Learning Pattern}, for integrating AI-driven personalization into automotive software architectures. These patterns enhance the application software layer by enabling adaptive, context-aware functions that personalize vehicle behavior while maintaining safety constraints.

We conclude the section with the exemplary mixed-criticality SWA shown in Fig.~\ref{fig:ExemplarySWA}, which incorporates these findings. While evidence emphasizes SWAs as \enquote{one of the most critical success factors for the design and development of complex software systems} \cite{SeedSet14}, Fig.~\ref{fig:ExemplarySWA} illustrates its strong relation to the HW layer beneath. The literature often discusses and explores architecture as a whole, considering the close relationship between HW and SW from the very beginning \cite{SeedSet12,SeedSet14,SeedSet21}. Lee and Wang \cite{SeedSet15} even highlight HW as the design driver for automotive architectures. While the SDV approach aims to decouple SW functions from HW \cite{SeedSet15}, the SWA and its requirements still depend on the HPC HW it runs on, and close co-design can help to solve this multiobjective problem \cite{SeedSet12}. 

The following section discusses threats to validity related to our SLR.

\section{Threats to Validity}\label{sec:TTV}
We evaluate our study's quality by discussing the three main threats to validity in secondary software engineering studies as identified by Ampatzoglou et al. \cite{TTV}: \textit{threats to study selection}, \textit{threats to data collection}, and \textit{threats to research validity}.

\subsection{Threats to Study Selection}
Our study omits snowballing due to the balanced F-measure of Scopus searches, as detailed in Section~\ref{sec:RP}. While some relevant papers were excluded due to access restrictions or language barriers, such as Chinese-language papers and SAE reports, the high precision of Scopus ensured a comprehensive seed set to answer our research question. We acknowledge potential publication bias and propose a search strategy for snowballing with our seed set as a basis to explore additional methodological and technical approaches in the outlook section. 
Additionally, industry insights may be outdated due to limited disclosure. Including gray literature or industry expert interviews could enhance understanding and identify current practices but conflicts with our focus on peer-reviewed studies. Concluding, we want to emphasize the potential threat by variability in evaluating loose inclusion criteria wordings as \enquote{clearly demonstrate} and \enquote{tend to answer}. 

\subsection{Threats to Data Collection}
We mitigated data extraction bias by using commonly agreed-upon data extraction criteria, with each study independently reviewed by two authors. Disagreements were resolved through discussion and consensus-building. Pre-piloting of the extraction criteria ensured a common understanding.  
As the study neither applies snowballing nor includes gray literature, it persists the threat of publication bias related to the data collected. To still ensure validity of the related study and reduce publication bias, we considered quality-related inclusion criteria (see Table~\ref{tab1}) as proposed by Kitchenham et al. \cite{kitchenhamguidelines}. Furthermore, the diversity of journals and conferences from which the included publications were drawn helps to mitigate the threat of publication bias.

\subsection{Threats to Research Validity}
To minimize research bias, the first four authors collaboratively designed and refined the review protocol, aligning with established guidelines. The first three authors specialize in automotive SW engineering, while the fourth focuses on AI integration into SWAs. The fifth and sixth authors -- experts in functional safety and SW engineering, respectively -- reviewed the protocol to further reduce bias. Disagreements on protocol, data extraction, and synthesis were resolved within the core group or escalated to the reviewers. 
To enable repeatability, all review discussions, protocol changes, and inclusion/exclusion decisions are documented and made publicly available on \href{https://github.com/AutomotiveArchitectures/ECSA-2025}{GitHub} and \href{https://doi.org/10.5281/zenodo.15093879}{Zenodo}.
This study focuses on automotive SWAs and may have limited applicability to other domains due to industry-specific constraints. Future research could explore cross-industry approaches, such as avionics systems (see IMA and ARINC standards \cite{SeedSet17}), or IT and cloud-based architectures, to identify transferable best practices.

\section{Conclusion and Outlook}\label{sec:CO}
In this paper, we systematically reviewed existing research on mixed-criticality SWAs for centralized HPC platforms in SDVs. To ensure methodological rigor and transparency in our analysis, we employed a structured research protocol grounded in established literature review practices. Through this protocol, we identified key functional domains, constraints, and enabling technologies that shape modern automotive SWAs. Furthermore, we extracted architectural patterns and design practices that support the integration of mixed-criticality requirements within centralized automotive SWAs. Our findings provide practitioner-oriented insights to aid SW architects and developers in designing HPC-based SWAs.

However, our study also revealed limitations and open challenges. The lack of publicly available industry data hinders validating recent advancements in automotive SWAs. Moreover, the transferability of best practices from other domains, such as avionics or cloud-based SWAs, requires further investigation.

Future work can build on the review protocol and expand the empirical findings with the hybrid search strategy \textit{Scopus + BS||FS} based on its consistent high F-measure values as analyzed by Mourão et al. \cite{hybrid1}. Our study delivers the seed set for parallel Backward Snowballing (BS) -- which increases recall by identifying references cited within a study -- and Forward Snowballing (FS) -- which improves precision by identifying newer studies citing a given publication \cite{hybrid1}. In addition, the study of cross-industry approaches could reveal transferable design principles for automotive applications. To capture the latest advances, the inclusion of gray literature and an empirical interview study with industry experts could uncover emerging approaches and technologies.

\begin{credits}
\subsubsection{\ackname} This research was partially supported by the German Federal Ministry of Education and Research in the project AutoDevSafeOps (01IS22087R) and the Incentive Fund of TUM Campus Heilbronn.

\subsubsection{\discintname}
The authors declare that they have no known competing financial interests or personal relationships that could have appeared to
influence the work reported in this paper.
\end{credits}
%
%
%

\begin{thebibliography}{10}
\providecommand{\url}[1]{\texttt{#1}}
\providecommand{\urlprefix}{URL }
\providecommand{\doi}[1]{https://doi.org/#1}

\bibitem{SeedSet11}
Akkaya, S., et~al.: A modular five-layered v-shaped architecture for autonomous
  vehicles. In: 2019 11th International Conference on Electrical and
  Electronics Engineering (ELECO). pp. 850--854. IEEE (2019),
  \url{https://doi.org/10.23919/ELECO47770.2019.8990402}

\bibitem{RW-SLRonSWArchOptiMeth}
Aleti, A., et~al.: Software architecture optimization methods: A systematic
  literature review. IEEE Transactions on Software Engineering  \textbf{39}(5),
   658--683 (2012), \url{https://doi.org/10.1109/TSE.2012.64}

\bibitem{TTV}
Ampatzoglou, A., et~al.: Identifying, categorizing and mitigating threats to
  validity in software engineering secondary studies. Information and software
  technology  \textbf{106},  201--230 (2019),
  \url{https://doi.org/10.1016/j.infsof.2018.10.006}

\bibitem{SeedSet18}
Askaripoor, H., et~al.: E/e architecture synthesis: Challenges and
  technologies. Electronics  \textbf{11}(4), ~518 (2022),
  \url{https://doi.org/10.3390/electronics11040518}

\bibitem{autosar-adaptive}
AUTOSAR: Autosar adaptive platform (2024), available at:
  \url{https://www.autosar.org/standards/adaptive-platform}. {Accessed:
  2025-03-07}

\bibitem{autosar-classic}
AUTOSAR: Autosar classic platform (2024), available at:
  \url{https://www.autosar.org/standards/classic-platform}. {Accessed:
  2025-03-07}

\bibitem{RW-SLRonBigData}
Avci, C., et~al.: Software architectures for big data: a systematic literature
  review. Big Data Analytics  \textbf{5}(1), ~5 (2020),
  \url{https://doi.org/10.1186/s41044-020-00045-1}

\bibitem{Bandur}
Bandur, V., et~al.: Making the case for centralized automotive e/e
  architectures. IEEE Transactions on Vehicular Technology  \textbf{70}(2),
  1230--1245 (2021), \url{https://doi.org/10.1109/TVT.2021.3054934}

\bibitem{RW-SLRonCoT}
Banijamali, A., et~al.: Software architectures of the convergence of cloud
  computing and the internet of things: A systematic literature review.
  Information and Software Technology  \textbf{122},  106271 (2020),
  \url{https://doi.org/10.1016/j.infsof.2020.106271}

\bibitem{ExtCom}
Bauer, T., et~al.: Reference architectures for automotive software. In:
  Reference Architectures for Critical Domains: Industrial Uses and Impacts,
  pp. 73--111. Springer (2022),
  \url{https://doi.org/10.1007/978-3-031-16957-1_5}

\bibitem{SeedSet3}
Bordoloi, U., et~al.: Autonomy-driven emerging directions in software-defined
  vehicles. In: 2023 Design, Automation \& Test in Europe Conference \&
  Exhibition (DATE). pp.~1--6. IEEE (2023),
  \url{https://doi.org/10.23919/DATE56975.2023.10136910}

\bibitem{SeedSet14}
Bucaioni, A., et~al.: Modelling centralised automotive e/e software
  architectures. Advanced Engineering Informatics  \textbf{59},  102289 (2024),
  \url{https://doi.org/10.1016/j.aei.2023.102289}

\bibitem{MCS}
Cinque, M., et~al.: Certify the uncertified: Towards assessment of
  virtualization for mixed-criticality in the automotive domain. In: 2022 52nd
  Annual IEEE/IFIP International Conference on Dependable Systems and Networks
  Workshops (DSN-W). pp. 8--11. IEEE (2022),
  \url{https://doi.org/10.1109/DSN-W54100.2022.00012}

\bibitem{SeedSet12}
Collin, A., et~al.: Autonomous driving systems hardware and software
  architecture exploration: optimizing latency and cost under safety
  constraints. Systems Engineering  \textbf{23}(3),  327--337 (2020),
  \url{https://doi.org/10.1002/sys.21528}

\bibitem{VSS}
COVESA: Vehicle signal specification (2025), available at:
  \url{https://covesa.github.io/vehicle_signal_specification/introduction/index.html}.
  {Accessed: 2025-03-07}

\bibitem{SeedSet21}
Druml, N., et~al.: Time-of-flight 3d imaging for mixed-critical systems. In:
  2015 IEEE 13th International Conference on Industrial Informatics (INDIN).
  pp. 1432--1437. IEEE (2015), \url{https://doi.org/10.1109/INDIN.2015.7281943}

\bibitem{SeedSet5}
El-Bayoumi, A.: An enhanced algorithm for memory systematic faults detection in
  multicore architectures suitable for mixed-critical automotive applications.
  International Journal of Safety and Security Engineering  \textbf{10}(4),
  467--474 (2020), \url{https://doi.org/10.18280/ijsse.100405}

\bibitem{SeedSet4}
Ferraro, D., et~al.: Time-sensitive autonomous architectures. Real-Time Systems
   \textbf{59}(4),  568--608 (2023),
  \url{https://doi.org/10.1007/s11241-023-09404-2}

\bibitem{RW-SMSAutoSWEng}
Haghighatkhah, A., et~al.: Automotive software engineering: A systematic
  mapping study. Journal of Systems and Software  \textbf{128},  25--55 (2017),
  \url{https://doi.org/10.1016/j.jss.2017.03.005}

\bibitem{SeedSet2}
Holstein, T., Wietzke, J.: Contradiction of separation through virtualization
  and inter virtual machine communication in automotive scenarios. In:
  Proceedings of the 2015 European Conference on Software Architecture
  Workshops. pp.~1--5 (2015), \url{https://doi.org/10.1145/2797433.2797437}

\bibitem{HarRed}
Kadry, H.M., et~al.: Electrical architecture and in-vehicle networking:
  Challenges and future trends. In: 2022 IEEE International Symposium on
  Circuits and Systems (ISCAS). pp. 1009--1013. IEEE (2022),
  \url{https://doi.org/10.1109/ISCAS48785.2022.9937481}

\bibitem{kitchenhamguidelines}
Kitchenham, B., et~al.: Guidelines for performing systematic literature reviews
  in software engineering. Tech. rep., Technical report, ver. 2.3 ebse (2007)

\bibitem{SeedSet20}
Kugele, S., et~al.: Architectural patterns for cross-domain personalised
  automotive functions. In: 2020 IEEE international conference on software
  architecture (ICSA). pp. 191--201. IEEE (2020),
  \url{https://doi.org/10.1109/ICSA47634.2020.00026}

\bibitem{SeedSet15}
Lee, J., Wang, L.: A method for designing and analyzing automotive software
  architecture: A case study for an autonomous electric vehicle. In: 2021
  International Conference on Computer Engineering and Artificial Intelligence
  (ICCEAI). pp. 20--26. IEEE (2021),
  \url{https://doi.org/10.1109/ICCEAI52939.2021.00004}

\bibitem{SeedSet19}
Lex, J., et~al.: Hyfar: A hypervisor-based fault tolerance approach for
  heterogeneous automotive real-time systems. Journal of Systems Architecture
  \textbf{156},  103263 (2024),
  \url{https://doi.org/10.1016/j.sysarc.2024.103263}

\bibitem{SeedSet16}
Li, Y., et~al.: Key technology and standardization route for new electronic and
  electrical architecture of intelligent and connected vehicles. In: 2023 3rd
  International Conference on Electrical Engineering and Control Science
  (IC2ECS). pp. 323--328. IEEE (2023),
  \url{https://doi.org/10.1109/IC2ECS60824.2023.10493525}

\bibitem{ParadigmShiftToZonal}
Lu, S., et~al.: A comparison of end-to-end architectures for connected
  vehicles. In: 2022 Fifth International Conference on Connected and Autonomous
  Driving. pp. 72--80. IEEE (2022),
  \url{https://doi.org/10.1109/MetroCAD56305.2022.00015}

\bibitem{Dijkstra}
Maier, J., Reuss, H.C.: Design of zonal e/e architectures in vehicles using a
  coupled approach of k-means clustering and dijkstra’s algorithm. Energies
  \textbf{16}(19), ~6884 (2023), \url{https://doi.org/10.3390/en16196884}

\bibitem{mauser2024centralization}
Mauser, L., Wagner, S.: Centralization potential of automotive e/e
  architectures. Journal of Systems and Software p. 112220 (2024),
  \url{https://doi.org/10.1016/j.jss.2024.112220}

\bibitem{MauserWS}
Mauser, L., et~al.: Methodical approach for centralization evaluation of modern
  automotive e/e architectures. In: European Conference on Software
  Architecture. pp. 165--179. Springer (2022),
  \url{https://doi.org/10.1007/978-3-031-36889-9_13}

\bibitem{hybrid1}
Mour{\~a}o, E., et~al.: On the performance of hybrid search strategies for
  systematic literature reviews in software engineering. Information and
  software technology  \textbf{123},  106294 (2020),
  \url{https://doi.org/10.1016/j.infsof.2020.106294}

\bibitem{SeedSet13}
Niedballa, D., Reuss, H.C.: Mpsoc-based platform for fail-operational control
  of an automated research vehicle. Journal of Tongji University (Natural
  Science)  \textbf{50}(S1),  151--155 (2024),
  \url{https://doi.org/10.11908/j.issn.0253-374x.23717}

\bibitem{Rayyan}
Ouzzani, M., et~al.: Rayyan—a web and mobile app for systematic reviews.
  Systematic reviews  \textbf{5},  1--10 (2016),
  \url{https://doi.org/10.1186/s13643-016-0384-4}

\bibitem{VolvoCICD}
Pelliccione, P., et~al.: Automotive architecture framework: The experience of
  volvo cars. Journal of systems architecture  \textbf{77},  83--100 (2017),
  \url{https://doi.org/10.1016/j.sysarc.2017.02.005}

\bibitem{SeedSet9}
P{\"u}llen, D., et~al.: A security process for the automotive service-oriented
  software architecture. IEEE Transactions on Vehicular Technology
  \textbf{73}(4),  5036--5053 (2023),
  \url{https://doi.org/10.1109/ETFA.2019.8868957}

\bibitem{SeedSet8}
Savithry, J., et~al.: Design of criticality-aware scheduling for advanced
  driver assistance systems. In: 2019 24th IEEE International Conference on
  Emerging Technologies and Factory Automation (ETFA). pp. 1407--1410. IEEE
  (2019), \url{https://doi.org/10.1109/ETFA.2019.8868957}

\bibitem{SeedSet17}
Sommer, S., et~al.: Race: A centralized platform computer based architecture
  for automotive applications. In: 2013 IEEE International Electric Vehicle
  Conference (IEVC). pp.~1--6. IEEE (2013),
  \url{https://doi.org/10.1109/IEVC.2013.6681152}

\bibitem{SeedSet10}
St{\"a}hle, H., et~al.: Towards the deployment of a centralized ict
  architecture in the automotive domain. In: 2013 2nd Mediterranean Conference
  on Embedded Computing. pp. 66--69. IEEE (2013),
  \url{https://doi.org/10.1109/MECO.2013.6601320}

\bibitem{100ECUs}
Staron, M.: Automotive software architectures. Springer (2021),
  \url{https://doi.org/10.1007/978-3-030-65939-4}

\bibitem{SeedSet6}
Sundharam, S.M., et~al.: Software architecture modeling of autosar-based
  multi-core mixed-critical electric powertrain controller. Modelling
  \textbf{2}(4),  706--727 (2021),
  \url{https://doi.org/10.3390/modelling2040038}

\bibitem{CASEdrivesZonal}
Surjekar, N.N., et~al.: A case study on migrating towards functionally safe
  zonal architecture using mbse. In: INCOSE International Symposium. vol.~33,
  pp. 1403--1417. Wiley Online Library (2023),
  \url{https://doi.org/10.1002/iis2.13089}

\bibitem{Vogelsang}
Vogelsang, A.: Feature dependencies in automotive software systems: Extent,
  awareness, and refactoring. Journal of Systems and Software  \textbf{160},
  110458 (2020), \url{https://doi.org/10.1016/j.jss.2019.110458}

\bibitem{SeedSet7}
Wallin, P., et~al.: Problems and their mitigation in system and software
  architecting. Information and Software Technology  \textbf{54}(7),  686--700
  (2012), \url{https://doi.org/10.1016/j.infsof.2012.01.004}

\bibitem{SDV}
Xie, G., et~al.: Recent advances and future trends for automotive functional
  safety design methodologies. IEEE Transactions on Industrial Informatics
  \textbf{16}(9),  5629--5642 (2020),
  \url{https://doi.org/10.1109/TII.2020.2978889}

\bibitem{Zerf}
Zerfowski, D., Lock, A.: Functional architecture and e/e-architecture--a
  challenge for the automotive industry. In: 19. Internationales Stuttgarter
  Symposium: Automobil-und Motorentechnik. pp. 909--920. Springer (2019),
  \url{https://doi.org/10.1007/978-3-658-25939-6_70}

\bibitem{SeedSet1}
Zhou, X., et~al.: Development of vehicle domain controller based on ethernet.
  In: Journal of Physics: Conference Series. vol.~1802, p. 022065. IOP
  Publishing (2021), \url{https://doi.org/10.1088/1742-6596/1802/2/022065}

\end{thebibliography}


\end{document}